\documentclass[%
 aip,
 jmp,%
 amsmath,amssymb,
reprint,%
]{revtex4-1}

\usepackage{graphicx}
\usepackage{dcolumn}
\usepackage{bm}
\usepackage{color}
\usepackage[mathlines]{lineno}

\begin{document}
\preprint{AIP/123-QED}

\title[Impedance Matched Absorptive Thermal Blocking Filters]{Impedance Matched Absorptive Thermal Blocking Filters}
\author{E.J. Wollack}
\email{edward.j.wollack@nasa.gov}
\affiliation{NASA Goddard Space Flight Center, Greenbelt, MD 20771}
\author{D.T. Chuss}
\affiliation{NASA Goddard Space Flight Center, Greenbelt, MD 20771}
\author{K. Rostem}
\affiliation{NASA Goddard Space Flight Center, Greenbelt, MD 20771}
\affiliation{Department of Physics and Astronomy, \\ The Johns Hopkins University, Baltimore, MD 21218}
\author{K. U-Yen}
\affiliation{NASA Goddard Space Flight Center, Greenbelt, MD 20771}
\date{\today}

\begin{abstract}
We have designed, fabricated and characterized absorptive thermal blocking filters for cryogenic microwave applications. The transmission line filter's input characteristic impedance is designed to match $50\,\Omega$  and its response has been validated from 0-to-50\,GHz. The observed return loss in the 0-to-20\,GHz design band is greater than $20\,$dB and shows graceful degradation with frequency. Design considerations and equations are provided that enable this approach to be scaled and modified for use in other applications.
\end{abstract}


\maketitle

\begin{quotation}
\end{quotation}

\section{\label{sec:introduction}Introduction}
Thermal blocking filters find wide use in cryogenic applications ranging from quantum computing to ultra-low-noise detectors. They can be used to provide the environmental isolation between cooled devices and the warmer temperature supporting bias and readout circuitry.  In particular, they are effective in rejecting thermal radiation, limiting radio frequency interference, providing a convenient means of heat sinking signal lines, and realizing a vacuum feedthrough.  In a microwave instrumentation setting a well-defined characteristic impedance, typically approximating a short or matched boundary condition, is desirable. From a radiometric perspective, such filter structures limit the available power by modifying the transmitted response and effectively reducing the photon density of states in the Planck distribution to a single dimension.

A variety of thermal blocking filter construction techniques and designs have been discussed in the literature. In the device's most basic form, a large shunt capacitor forms a single-pole low-pass-filter.~\cite{Bladh2003} More generally, multiple low-pass lumped element stages can be combined in series to produce compact and broadband non-dissipative filter structures.~\cite{Vion1995,Sueur2006,UYen2008} The challenges presented by these implementations include controlling inter-stage isolation and spurious transmission resonances, limiting the filter's total shunt capacitance, and achieving adequate control over circuit parameters as a function of temperature.~\cite{Brown2012} Dissipative solutions based on distributed lossy microwave structures~\cite{Schiffres1964,Martinis1987,Zorin1995,Fukushima1997,Leong2002,Lukashenko2008} can be used to achieve a broadband low-pass transmission response. More recent efforts have strived to retain these desirable properties while providing a well defined impedance match. Specific examples in this class include coaxial-lines~\cite{Milliken2007} and strip-line~\cite{Santavicca2008,Slichter2009} powder filters. 

In this work, simple matched filter designs based on easily realized absorptive dielectric transmission lines are improved upon, and the resulting performance is described in detail. The filter's response is calculable, repeatable under cryogenic cycling, and is capable of providing an intrinsically broadband matched impedance termination. In Section~\ref{sec:design}, practical design considerations and the governing equations for the filter's operation are described in detail. The fabrication and test of the representative filters are summarized in Section~\ref{sec:fabrication}.

\section{\label{sec:design}Design Method and Considerations}
\begin{figure}[!h]
	\includegraphics[width=3.3 in]{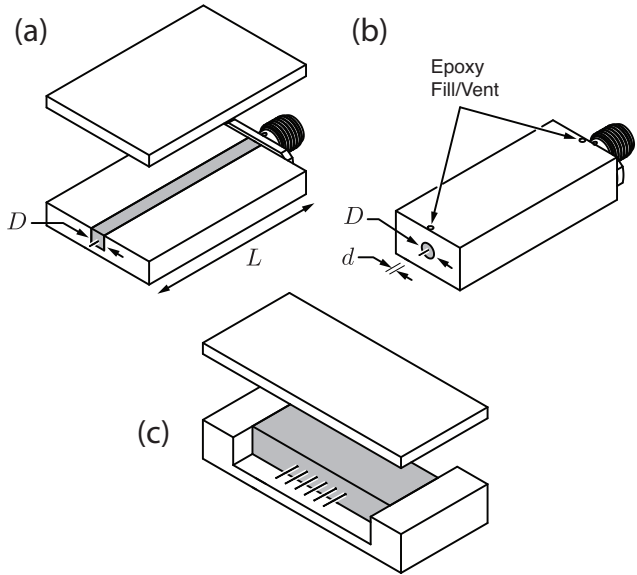}
	\vspace{0pt}
	\caption{Exploded view of thermal blocking filter geometry.  Package configurations for the following filter configurations are shown: a) milled channel, b) drilled hole, and c) multi-wire. For configuration a) the channel is filled with loaded epoxy from each end of the channel to the center to reduce risk of voids in the mixture near the connector interfaces. The fill/vent holes indicated in b) are filled with a metal insert after potting. Configuration c) can be used in conjunction with a controlled impedance fan-out board. For clarity, the connectors, signal termination, and fasteners on the left side of the sketch are not depicted.}
	\label{fig:filter_geometry}
\end{figure}

In order to minimize reflections in an absorptive transmission line filter, the modal symmetry, geometric overlap, and impedance of the lines and interface discontinuities need to be appropriately matched. From this perspective, with appropriate attention to detail, a tube with a centered wire and lossy dielectric lends itself to fulfill the desired low-pass filter functions. The structure can be easily fabricated via direct machining with readily available materials and can provide reliable operation in a cryogenic environment. See Figure~\ref{fig:filter_geometry} for a sketch of the filter geometries explored and characterized here. 

A simplified microwave model that assumes a constant dielectric function and retains the qualitative response of the device is presented. The first step is to select a realizable geometry and lossy medium that present a characteristic impedance equal to that of the measurement system. The characteristic impedance of a single wire in a square enclosure can be expressed as a perturbative expansion on a coaxial line~\cite{ITT1975} or in closed form through the introduction of a shielding factor~\cite{Wheeler1950,Riblet1983} as follows:
\begin{equation}
	\hat{Z_o} = { {\eta_o \over 2 \pi}  }  \cdot {\left( \mu_r   \over \epsilon_r \right)^{1/2} } \cdot { \ln \left({ D \over d}  \cdot s\left({ D / d} \right)\right)} ,
	\label{equ:impedance}
\end{equation}
where $\mu_r$ and $\epsilon_r$ are the relative permeability and permittivity of the media between the inner and outer conductors in the filter respectively, and $\eta_o = (\mu_o / \epsilon_o)^{1/2} \simeq 377 \, \Omega$\,per square is the wave impedance of free space. Here $D$ is the width (height) of the channel and $d$ is the center conductor outer diameter. The shielding factor, $s(D/d)$,  is a weak monotonically increasing function for this topology. Numerical simulations for a circular wire in a square enclosure reveal $s(1.1) \approx 1.06422$ and rapidly approaches a limiting asymptote, $s(\infty) \rightarrow 1.07871$, for $D/d > 3$. 

Similarly, the impedance of a wire centered between ground parallel planes can be computed from the above expression by adopting a shielding factor of $s(D/d)=4/\pi$. For a coaxial transmission line, the shielding factor, $s(D/d)$, is unity and $D$ is taken as the inner diameter of the outer conductor wall. The sensitivity of the impedance to the conductor centering in these configurations can be analytically treated by considering the structure as an asymmetric line~\cite{ITT1975}  or eccentric line~\cite{Moreno1944}, respectively. 

In practice, the lossy transmission line section has sufficient length to suppress coherent reflections (standing waves) between interfaces at the ends of the line. As a result, the power transmission response follows Beer's law,~\cite{Bohren2004}
\begin{equation}
	T \simeq \left( 1- R \right)^2 \cdot \exp( - \alpha L ) ,
	\label{equation:filtertransmission}
\end{equation}
where $\alpha \equiv 4 \pi \kappa / \lambda$ is the filter's power absorption coefficient, $\lambda$ is the radiation wavelength, and $L$ is the length of the lossy dielectric line. Here $R = {[(n-n_o)^2+\kappa^2] / [(n+n_o)^2+\kappa^2]}$ is the reflected power at the interface between the lossless and lossy dielectric media characterized by complex indices of refraction, $n_o$ and $\hat{n} = n + i \kappa$, respectively. Solving for the real and imaginary component of $\hat{n}$ in terms of the relative dielectric permittivity, $\epsilon_r = \epsilon_r' + i \cdot \epsilon_r'' = (n^2-\kappa^2) + 2 i n \kappa$, one finds:
\begin{equation}
	n          =  \sqrt{ \left( \epsilon_r'^2 + \epsilon_r''^2 \right)^{1 \over 2} +  \epsilon_r' \over 2 }, \,\,\,\,
	\kappa =  \sqrt{ \left( \epsilon_r'^2 + \epsilon_r''^2 \right)^{1 \over 2} -  \epsilon_r' \over 2 }.
\end{equation}
This enables calculation of the frequency dependance of the filter transmission function. Calculation of the filter's effective bandwidth is presented in the Appendix. More generally, expressions for the propagation constant in the presence of magnetic materials~\cite{VonHipple1954} can be employed and can be used to achieve further reduction in component volume and transmission line density if desired. 

To the extent the dielectric function can be treated as a constant over the range of interest, the ratio of the power attenuation coefficient over the frequency is a constant for a given material. This property can be used to define the filter's roll off by adjusting the length of the lossy dielectric section to present the desired attenuation above a specified frequency. For manufacturing simplicity the mismatch arising at the lossless-lossy media interfaces is uncompensated in the design outlined. As a result, the power reflection without compensation is limited to $R > {\kappa^2 / (4n^2+\kappa^2)} \approx (\epsilon_r''/4\epsilon_r')^2$ for $n = n_o$ and a dielectric loss tangent, $\tan\delta \equiv \epsilon_r''/\epsilon_r' < 1$. A fundamental trade exists between minimizing reflectance and the transmission response achievable within a given line length. 

In practice we rely on a more refined transmission line model and finite element simulations performed in HFSS (High Frequency Structure Simulator) to incorporate the slight frequency dependance~\cite{Wollack2008} of the dielectric media and sensitivity to variations from the nominal design. For the filter realizations and tolerances described here, the calculations performed by the differing approaches are in agreement.

\section{\label{sec:fabrication}Filter Fabrication and Test}
In the matched filter implementations explored, a stainless-steel wire ($\approx0.2$~mm outer diameter) serves as the center conductor and a metal housing forms the metallization for line's outer conductor. The stainless steel wire diameter was selected to be compatible and make a reliable light press fit into the SMA connector's rear socket (Southwest Microwave, Tempe AZ; model number 214-537SF; nominal pin inner diameter  $\approx 0.225$\,mm, pin insertion depth $\approx 2$\,mm). This connector supports a single transverse electromagnetic mode through $\sim27$\,GHz, and the filter provides sufficient attenuation to prevent spurious resonances in the structure and degradation of the impedance response~\cite{Denny1968}. The measured total DC resistance of the connectorized structures (e.g., $<1.4\,\Omega$ for the square channel filter described) are observed to be within $4\%$ of the value computed for 304 stainless and the center conductor line geometry. This can be reduced by employing a copper clad stainless magnet wire or a superconducting center conductor if compatible with the desired end application. 

A dielectric mixture comprised of epoxy (Ellsworth Adhesives, Germantown, WI; Emerson-Cumming 2850 FT with Catalyst 23LV) loaded with a volume filling fraction $\simeq0.3$ of stainless-steel powder~\cite{Wollack2008} is used to fill the volume between the center and outer conductor forming a controlled impedance transmission line. The dielectric permittivity and magnetic permeability of this loading are $\epsilon_r \simeq 10.6+2.1i$ and  $\mu_r \approx 1$, respectively. This choice represents the highest permittivity mixture that has flow characteristics compatible with filling the desired volume and minimizes the envelope of the filter package. Using Equation~(\ref{equ:impedance}) to achieve an impedance match to a $50\,\Omega$ line, a $D/d \simeq 15$ ratio should be targeted when using this dielectric mixture. Reflections at the $-26\,$dB level are anticipated in evaluating the reflectance in Equation~(\ref{equation:filtertransmission}). The ratio of power attenuation coefficient over the operating frequency in gigahertz, $\nu_{\rm GHz}$, is $\alpha/\nu_{\rm GHz} \simeq 0.058 \pm 0.005$\,dB/mm/GHz, where the uncertainty is bounded by the observed variation in dielectric mixture's loss. The real component of the dielectric permittivity as a function of the volume filling fraction has a lower impact on the overall filter response.

The metal housings for these filters are realized by either drilling a circular hole through a block or milling a square channel in a split-block as indicated in Figure~\ref{fig:filter_geometry}. The primary objective in potting the filter is to achieve a homogenous dielectric mixture free of voids between the transmission line's center and outer conductors. This goal can be readily achieved for the square channel split-block filter configuration. The top surface of the milled block is temporarily masked with Kapton tape, the channel filled with epoxy, cured, and subsequently lapped smooth  to realize a low impedance metal-to-metal contact between the flat surfaces of the body and the cover. An average surface roughness, $R_a \le 0.2 \, \mu$m, is specified for the finish at all mating surfaces to achieve the desired RF seal.

In the drilled filter housing approach, a 0.98\,mm diameter hole approximately 3 diameters in length,  is made near each end of the tube for filling with epoxy and venting gas from the volume. The SMA connectors are installed in order to locate the transmission line center conductor during potting. This arrangement precisely defines the desired volume and allows the epoxy mixture to be injected into the filter with a syringe without entrapping bubbles. Once the cavity is filled with epoxy, the fill and vent holes are plugged with a 19\,gauge wire (0.91\,mm outer diameter) to be flush with the inner tube wall. This defines an electrically short lossy line section with $\simeq5 \, \Omega$ characteristic impedance which is small compared to the signal line $Z_o$. As a result, the arrangement has a negligible influence on the RF match at the SMA ports. 

\begin{figure}[!h]
	\includegraphics[width=3.5 in]{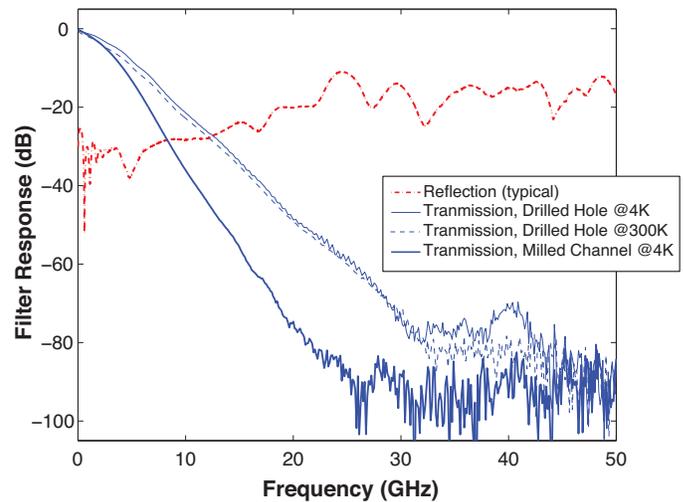}
	\vspace{0pt}
	\caption{Summary of the measured matched blocking filter responses. The bold-dashed (red) line indicates the observed reflectance at 300\,K for the milled channel filter and is representative for devices realized by the construction techniques described here. The thin-solid and -dashed (blue) lines show the change in the drilled hole filter transmission upon cooling. The bold-solid (blue) trace indicates the transmission response of a milled split-block channel filter. The difference between the drilled and milled configuration responses arises from the differing transmission line lengths and is consistent with $\alpha/\nu_{\rm GHz} \approx 0.058$\,dB/mm/GHz derived from measurement of the epoxy mixture coupons. The measured filter half-power response for the milled and drilled structures were 1.8 and 2.4\,GHz respectively. The noise floor of the VNA calibration was $>80\,$dB and degraded slightly upon cooling the test configuration.}
\label{fig:BlockingFilter}
\end{figure}

The residual loaded epoxy was used to make a pair of witness samples in WR28.0 waveguide of differing physical lengths.~\cite{Wollack2008} After curing the witness samples, the scattering parameters were measured with an Agilent PNA-X Vector Network Analyzer (VNA) from 26-40 GHz with a TRL (Thru-Reflect-Line) calibration. A two-line matrix technique~\cite{Janezic1999} was used to extract and verify the electromagnetic properties of the mixture employed in each filter structure. The scattering parameters of the filters were measured at room temperature and $\sim4$\,K. For these measurements the VNA was calibrated with a SOLT (Short-Open-Line-Thru) calibration in 2.4\,mm $50\,\Omega$ lines where the reference planes reside at the test device's SMA connectors. The parameters for the filters characterized are described in Table~\ref{tab:filter_params} and the response summarized in Figure~\ref{fig:BlockingFilter}. The measurements are consistent with a transmission line model~\cite{Pozar2004} calculation for the as built geometry and fabrication tolerances.
\begin{table}[!h]
	\centering
	\begin{tabular}{|r|r|r|r|c|} \hline \hline
                 Filter Housing                     &  Milled                 &  Drilled      &  6-Line      &                  \\
                 Construction                       &  Channel             &  Hole         &  Channel   &  Units        \\ \hline
	         Inner Conductor, $d$         &  0.200                  &  0.200       &  0.16         &   mm         \\ 
	         Outer Conductor, $D$        &  3.2                     &    3.0          &  36            &   mm         \\ 
	         Filter Length, $L$               &  62.5                   &  38.9          &  13            &   mm         \\ 
	         Shielding Factor, $s$         & $\simeq1.079$    & $1$            & $4/\pi$      &    -             \\ \hline \hline
	\end{tabular}
  	\caption{Thermal Blocking Filter Geometric Parameters}
	\label{tab:filter_params}
\end{table}

\begin{figure}[!h]
	\includegraphics[width=3.3 in]{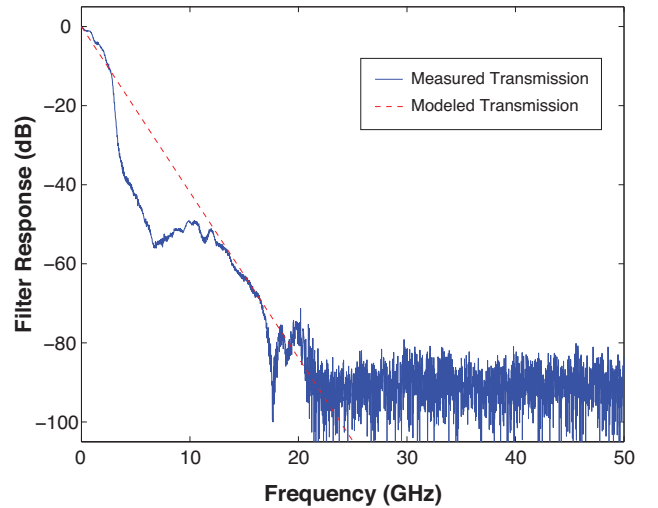}
	\vspace{0pt}
	\caption{Observed Multi-Line Filter Transmission Response (solid blue line). This prototype device was intended for use at low frequencies and no attempt has been made to maintain a broadband match through the package to the test connector interface. The dashed (red) line indicates the computed transmission for a 13\,mm transmission line filter section with $\alpha/\nu_{\rm GHz} \simeq 0.33$\,dB/mm/GHz. The presence of a dielectric permittivity with $\epsilon_r \simeq 16+16i$ is inferred from the transmission spectrum envelope for this realization of the filter structure.  The filter's measured half-power response frequency is $\sim1.1\,$GHz and the noise floor of the VNA calibration is approximately $85\,$dB. }
\label{fig:BlockingFilter_nline}
\end{figure}

We have also realized variations on the fabrication scheme described for filtering arrays of multiple signal lines in a wide channel with an alternative dielectric mixture. For the six-line filter example depicted in Figure~\ref{fig:filter_geometry}, PTFE shims were used to temporarily hold the conductors in place within a milled channel. The steel-powder was then poured to fill the filter volume. Once assembled, a $\sim2\,$~centipoise cyanoacrylate (i.e., ``low-viscosity superglue") was wicked through the powder and cured to realize the lossy dielectric insert. This mixture formulation's relatively large permittivity~\cite{Wollack2008}, $\epsilon_r \sim 20+20i$, correspondingly shrinks the dimensions required to maintain a given level of mode conversion, increases the tolerances to control the impedance, and increases the line attenuation ($\alpha/\nu_{\rm GHz} \simeq 0.37 \pm 0.04$\,dB/mm/GHz). In principle, the permittivity can be further increased by employing a disordered alloy with a higher conductivity than stainless steel or by incorporation of magnetic materials in the mixture. The mixture's relatively large dielectric loss tangent tends to degrades the filter's input impedance (e.g., a return loss of $\sim14$\,dB is computed for the formulation described), however, it enables a $>24\,$dB inter-line isolation to be achieved between adjacent lines. Use of every other line as a grounded guard line or terminating in the system's characteristic impedance would enable a modest improvement in isolation over this level; however, this reduces the available density of signal lines and may provide undesired signal return paths for low-noise applications. After removal of the PTFE shims, the top surface of the filter was lapped smooth for electrical contact between the body and cover. In this example, the total channel width is $36$\,mm with an interline spacing = 2.5\,mm was implemented. A summary of the measured filter response is presented in Figure~\ref{fig:BlockingFilter_nline}.

\vspace{0pt}
\section{\label{sec:Conclusion}Conclusion}
The electrical design, fabrication, and characterization of thermal blocking filters based upon a lossy transmission line are described. These impedance matched structures have been realized in circular and square tube geometries with low manufacturing complexity and terminated with readily available commercial connectors. A return loss $>20$dB from 0-to-20~GHz is observed. A prototype wide slot filter configuration for use with a multi-line fanout board is also presented. The circuit concepts described are flexible and can be readily adapted to meet a host of microwave metrology needs at cryogenic temperatures.  

\vspace{0pt}
\section*{Acknowledgements}
The authors gratefully acknowledge financial support from the NASA ROSES/APRA program and thank F. Colazo, A.J. Goulette, and M. Turvey for their contributions to the development of these structures.

\vspace{0pt}
\section*{Appendix: Filtered 1D Blackbody Radiation}

Assume that the blocking filter is optically thick, matched, and at a physical temperature small compared to the thermodynamic temperature, $T$, of the blackbody source to be rejected. In this limit, the thermal power coupled from a one-dimensional Planckian spectrum through a single-mode filter can be expressed as follows:
\begin{equation}
	P = \int_0^\infty {h\nu  \over e^{h \nu / k_b T} -1} \, \cdot F(\nu) \, d\nu.
\end{equation}  
Recall that the filter in question is matched and the filter response described by Equation~\ref{equation:filtertransmission} is well approximated by $F(\nu) \simeq \exp(-4 \pi \kappa L / \lambda) = \exp(-4 \pi \kappa L \nu / c)$, and an analytical evaluation of the desired expression can be derived. For convenience we define a dimensionless constant, $\beta \equiv 4 \pi \kappa L k_b T/ h c $, change variables to $x = h\nu/k_bT$, and find,

\begin{equation}
	P = {(k_b T)^2 \over h} \int_0^\infty {x e^{-(1+\beta) \cdot x} \over 1 - e^{-x} } \, dx .
\end{equation}   
By expanding the denominator as a geometric series,
\begin{equation}
	P = {(k_b T)^2 \over h} \int_0^\infty x e^{-(1+\beta) \cdot x}  \sum_{n=0}^\infty e^{-n x} \, dx ,
\end{equation}    
interchanging the order of integration and the summation and changing variables to $\zeta = (1+ \beta+n) \cdot x $ yields,
\begin{equation}
	P = {(k_b T)^2 \over h} \sum_{n=0}^\infty   {1 \over (1+\beta+n)^2 }   \int_0^\infty {\zeta e^{-\zeta}  } \, d\zeta .
\end{equation}    
Noting that the sum over $n$ gives the trigamma function~\cite{Ogreid2001} $\psi'_1\left( 1+\beta \right)$ and the integral is $\Gamma(2) = 1$, enables the coupled power to be expressed as follows: 
\begin{equation}
	P = {(k_b T)^2 \over h} \cdot \psi'_1\left( 1+\beta \right),
	\label{1D_power} 
\end{equation}    
where the trigamma function can be evaluated numerically or via its recursion relation, $\psi'_1(1+\beta)= \psi'_1(\beta)-1/\beta^2$, for $\beta>0$. With a lossless line, $\psi'_1(1) = \pi^2/6$, and Equation~\ref{1D_power} gives the physically anticipated power emitted by a 1D blackbody spectrum.~\cite{Landsberg1989,Menon1998}  Recalling $\beta$ is a dimensionless measure of the filter's roll-off frequency over the blackbody's characteristic response, for the thermal blocker design to be effective, $\beta\gg1$. Evaluating $\psi'_1(\beta) \rightarrow 1/\beta$ in its asymptotic limit we obtain an approximate form for the residual coupled power encountered in practical applications,
\begin{equation}
	P \simeq {(k_b T)^2 \over h \beta} =  k_b T \cdot {c \over 4 \pi \kappa L} = k_b T \cdot \Delta\nu_{\rm eff},
\end{equation}    
where $\Delta\nu_{\rm eff}  \equiv c / 4 \pi \kappa L \simeq  \nu_{0.5} / \ln(2)$ is the effective radiometric signal bandwidth and $\nu_{0.5}$ is frequency of the thermal blocking filter's  half-power response. The product, $\Delta\nu_{\rm eff} \cdot L \simeq 80 {\rm \,GHz \cdot mm}$, is estimated for a matched filter with $\epsilon_r \simeq 10.6+2.1i$ and $\mu_r \approx 1$ when used in the manner described.

\vspace{0pt}

\begin{thebibliography}{27}%
\makeatletter
\providecommand \@ifxundefined [1]{%
 \@ifx{#1\undefined}
}%
\providecommand \@ifnum [1]{%
 \ifnum #1\expandafter \@firstoftwo
 \else \expandafter \@secondoftwo
 \fi
}%
\providecommand \@ifx [1]{%
 \ifx #1\expandafter \@firstoftwo
 \else \expandafter \@secondoftwo
 \fi
}%
\providecommand \natexlab [1]{#1}%
\providecommand \enquote  [1]{``#1''}%
\providecommand \bibnamefont  [1]{#1}%
\providecommand \bibfnamefont [1]{#1}%
\providecommand \citenamefont [1]{#1}%
\providecommand \href@noop [0]{\@secondoftwo}%
\providecommand \href [0]{\begingroup \@sanitize@url \@href}%
\providecommand \@href[1]{\@@startlink{#1}\@@href}%
\providecommand \@@href[1]{\endgroup#1\@@endlink}%
\providecommand \@sanitize@url [0]{\catcode `\\12\catcode `\$12\catcode
  `\&12\catcode `\#12\catcode `\^12\catcode `\_12\catcode `\%12\relax}%
\providecommand \@@startlink[1]{}%
\providecommand \@@endlink[0]{}%
\providecommand \url  [0]{\begingroup\@sanitize@url \@url }%
\providecommand \@url [1]{\endgroup\@href {#1}{\urlprefix }}%
\providecommand \urlprefix  [0]{URL }%
\providecommand \Eprint [0]{\href }%
\providecommand \doibase [0]{http://dx.doi.org/}%
\providecommand \selectlanguage [0]{\@gobble}%
\providecommand \bibinfo  [0]{\@secondoftwo}%
\providecommand \bibfield  [0]{\@secondoftwo}%
\providecommand \translation [1]{[#1]}%
\providecommand \BibitemOpen [0]{}%
\providecommand \bibitemStop [0]{}%
\providecommand \bibitemNoStop [0]{.\EOS\space}%
\providecommand \EOS [0]{\spacefactor3000\relax}%
\providecommand \BibitemShut  [1]{\csname bibitem#1\endcsname}%
\let\auto@bib@innerbib\@empty
\bibitem [{\citenamefont {Bladh}\ \emph {et~al.}(2003)\citenamefont {Bladh},
  \citenamefont {Gunnarsson}, \citenamefont {Hurfeld}, \citenamefont {Devi},
  \citenamefont {Kristoffersson}, \citenamefont {Smalander}, \citenamefont
  {Pehrson}, \citenamefont {Claeson}, \citenamefont {Delsing},\ and\
  \citenamefont {Taslakov}}]{Bladh2003}%
  \BibitemOpen
  \bibfield  {author} {\bibinfo {author} {\bibfnamefont {K.}~\bibnamefont
  {Bladh}}, \bibinfo {author} {\bibfnamefont {D.}~\bibnamefont {Gunnarsson}},
  \bibinfo {author} {\bibfnamefont {E.}~\bibnamefont {Hurfeld}}, \bibinfo
  {author} {\bibfnamefont {S.}~\bibnamefont {Devi}}, \bibinfo {author}
  {\bibfnamefont {C.}~\bibnamefont {Kristoffersson}}, \bibinfo {author}
  {\bibfnamefont {B.}~\bibnamefont {Smalander}}, \bibinfo {author}
  {\bibfnamefont {S.}~\bibnamefont {Pehrson}}, \bibinfo {author} {\bibfnamefont
  {T.}~\bibnamefont {Claeson}}, \bibinfo {author} {\bibfnamefont
  {P.}~\bibnamefont {Delsing}}, \ and\ \bibinfo {author} {\bibfnamefont
  {M.}~\bibnamefont {Taslakov}},\ }\bibfield  {title} {\enquote {\bibinfo
  {title} {Comparison of cryogenic filters for use in single electronics
  experiments},}\ }\href {\doibase 10.1063/1.1540721} {\bibfield  {journal}
  {\bibinfo  {journal} {Review of Scientific Instruments}\ }\textbf {\bibinfo
  {volume} {74}},\ \bibinfo {pages} {1323 -- 1327} (\bibinfo {year}
  {2003})}\BibitemShut {NoStop}%
\bibitem [{\citenamefont {Vion}\ \emph {et~al.}(1995)\citenamefont {Vion},
  \citenamefont {Orfila}, \citenamefont {Joyez}, \citenamefont {Esteve},\ and\
  \citenamefont {Devoret}}]{Vion1995}%
  \BibitemOpen
  \bibfield  {author} {\bibinfo {author} {\bibfnamefont {D.}~\bibnamefont
  {Vion}}, \bibinfo {author} {\bibfnamefont {P.~F.}\ \bibnamefont {Orfila}},
  \bibinfo {author} {\bibfnamefont {P.}~\bibnamefont {Joyez}}, \bibinfo
  {author} {\bibfnamefont {D.}~\bibnamefont {Esteve}}, \ and\ \bibinfo {author}
  {\bibfnamefont {M.~H.}\ \bibnamefont {Devoret}},\ }\bibfield  {title}
  {\enquote {\bibinfo {title} {Miniature electrical filters for single electron
  devices},}\ }\href {\doibase 10.1063/1.358781} {\bibfield  {journal}
  {\bibinfo  {journal} {Journal of Applied Physics}\ }\textbf {\bibinfo
  {volume} {77}},\ \bibinfo {pages} {2519 -- 2524} (\bibinfo {year}
  {1995})}\BibitemShut {NoStop}%
\bibitem [{\citenamefont {le~Sueur}\ and\ \citenamefont
  {Joyez}(2006)}]{Sueur2006}%
  \BibitemOpen
  \bibfield  {author} {\bibinfo {author} {\bibfnamefont {H.}~\bibnamefont
  {le~Sueur}}\ and\ \bibinfo {author} {\bibfnamefont {P.}~\bibnamefont
  {Joyez}},\ }\bibfield  {title} {\enquote {\bibinfo {title} {Microfabricated
  electromagnetic filters for millikelvin experiments},}\ }\href {\doibase
  10.1063/1.2370744} {\bibfield  {journal} {\bibinfo  {journal} {Review of
  Scientific Instruments}\ }\textbf {\bibinfo {volume} {77}},\ \bibinfo {pages}
  {115102} (\bibinfo {year} {2006})}\BibitemShut {NoStop}%
\bibitem [{\citenamefont {U-yen}\ and\ \citenamefont
  {Wollack}(2008)}]{UYen2008}%
  \BibitemOpen
  \bibfield  {author} {\bibinfo {author} {\bibfnamefont {K.}~\bibnamefont
  {U-yen}}\ and\ \bibinfo {author} {\bibfnamefont {E.~J.}\ \bibnamefont
  {Wollack}},\ }\bibfield  {title} {\enquote {\bibinfo {title} {Compact planar
  microwave blocking filter},}\ }in\ \href {\doibase 10.1109/EUMC.2008.4751534}
  {\emph {\bibinfo {booktitle} {Microwave Conference, 2008. EuMC 2008. 38th
  European}}}\ (\bibinfo {year} {2008})\ pp.\ \bibinfo {pages} {642 --
  645}\BibitemShut {NoStop}%
\bibitem [{\citenamefont {Brown}\ \emph {et~al.}(2013)\citenamefont {Brown},
  \citenamefont {Chervenak}, \citenamefont {Chuss}, \citenamefont {Mikula},
  \citenamefont {Ray}, \citenamefont {Rostem}, \citenamefont {U-yen},
  \citenamefont {Wassell},\ and\ \citenamefont {Wollack}}]{Brown2012}%
  \BibitemOpen
  \bibfield  {author} {\bibinfo {author} {\bibfnamefont {A.}~\bibnamefont
  {Brown}}, \bibinfo {author} {\bibfnamefont {J.}~\bibnamefont {Chervenak}},
  \bibinfo {author} {\bibfnamefont {D.}~\bibnamefont {Chuss}}, \bibinfo
  {author} {\bibfnamefont {V.}~\bibnamefont {Mikula}}, \bibinfo {author}
  {\bibfnamefont {C.}~\bibnamefont {Ray}}, \bibinfo {author} {\bibfnamefont
  {K.}~\bibnamefont {Rostem}}, \bibinfo {author} {\bibfnamefont
  {K.}~\bibnamefont {U-yen}}, \bibinfo {author} {\bibfnamefont
  {E.}~\bibnamefont {Wassell}}, \ and\ \bibinfo {author} {\bibfnamefont
  {E.}~\bibnamefont {Wollack}},\ }\bibfield  {title} {\enquote {\bibinfo
  {title} {Fabrication of compact superconducting lowpass filters for
  ultrasensitive detectors},}\ }\href {\doibase 10.1109/TASC.2012.2231135}
  {\bibfield  {journal} {\bibinfo  {journal} {Applied Superconductivity, IEEE
  Transactions on}\ }\textbf {\bibinfo {volume} {23}},\ \bibinfo {pages}
  {2300204 -- 2300204} (\bibinfo {year} {2013})}\BibitemShut {NoStop}%
\bibitem [{\citenamefont {Schiffres}(1964)}]{Schiffres1964}%
  \BibitemOpen
  \bibfield  {author} {\bibinfo {author} {\bibfnamefont {P.}~\bibnamefont
  {Schiffres}},\ }\bibfield  {title} {\enquote {\bibinfo {title} {A dissipative
  coaxial rfi filter},}\ }\href {\doibase 10.1109/TEMC.1964.4307330} {\bibfield
   {journal} {\bibinfo  {journal} {Electromagnetic Compatibility, IEEE
  Transactions on}\ }\textbf {\bibinfo {volume} {6}},\ \bibinfo {pages} {55 --
  61} (\bibinfo {year} {1964})}\BibitemShut {NoStop}%
\bibitem [{\citenamefont {Martinis}, \citenamefont {Devoret},\ and\
  \citenamefont {Clarke}(1987)}]{Martinis1987}%
  \BibitemOpen
  \bibfield  {author} {\bibinfo {author} {\bibfnamefont {J.~M.}\ \bibnamefont
  {Martinis}}, \bibinfo {author} {\bibfnamefont {M.~H.}\ \bibnamefont
  {Devoret}}, \ and\ \bibinfo {author} {\bibfnamefont {J.}~\bibnamefont
  {Clarke}},\ }\bibfield  {title} {\enquote {\bibinfo {title} {Experimental
  tests for the quantum behavior of a macroscopic degree of freedom: The phase
  difference across a josephson junction},}\ }\href {\doibase
  10.1103/PhysRevB.35.4682} {\bibfield  {journal} {\bibinfo  {journal} {Phys.
  Rev. B}\ }\textbf {\bibinfo {volume} {35}},\ \bibinfo {pages} {4682 -- 4698}
  (\bibinfo {year} {1987})}\BibitemShut {NoStop}%
\bibitem [{\citenamefont {Zorin}(1995)}]{Zorin1995}%
  \BibitemOpen
  \bibfield  {author} {\bibinfo {author} {\bibfnamefont {A.~B.}\ \bibnamefont
  {Zorin}},\ }\bibfield  {title} {\enquote {\bibinfo {title} {The thermocoax
  cable as the microwave frequency filter for single electron circuits},}\
  }\href {\doibase 10.1063/1.1145385} {\bibfield  {journal} {\bibinfo
  {journal} {Review of Scientific Instruments}\ }\textbf {\bibinfo {volume}
  {66}},\ \bibinfo {pages} {4296 -- 4300} (\bibinfo {year} {1995})}\BibitemShut
  {NoStop}%
\bibitem [{\citenamefont {Fukushima}\ \emph {et~al.}(1997)\citenamefont
  {Fukushima}, \citenamefont {Sato}, \citenamefont {Iwasa}, \citenamefont
  {Nakamura}, \citenamefont {Komatsuzaki},\ and\ \citenamefont
  {Sakamoto}}]{Fukushima1997}%
  \BibitemOpen
  \bibfield  {author} {\bibinfo {author} {\bibfnamefont {A.}~\bibnamefont
  {Fukushima}}, \bibinfo {author} {\bibfnamefont {A.}~\bibnamefont {Sato}},
  \bibinfo {author} {\bibfnamefont {A.}~\bibnamefont {Iwasa}}, \bibinfo
  {author} {\bibfnamefont {Y.}~\bibnamefont {Nakamura}}, \bibinfo {author}
  {\bibfnamefont {T.}~\bibnamefont {Komatsuzaki}}, \ and\ \bibinfo {author}
  {\bibfnamefont {Y.}~\bibnamefont {Sakamoto}},\ }\bibfield  {title} {\enquote
  {\bibinfo {title} {Attenuation of microwave filters for single-electron
  tunneling experiments},}\ }\href {\doibase 10.1109/19.571834} {\bibfield
  {journal} {\bibinfo  {journal} {Instrumentation and Measurement, IEEE
  Transactions on}\ }\textbf {\bibinfo {volume} {46}},\ \bibinfo {pages} {289
  -- 293} (\bibinfo {year} {1997})}\BibitemShut {NoStop}%
\bibitem [{\citenamefont {Leong}\ \emph {et~al.}(2002)\citenamefont {Leong},
  \citenamefont {Bhatia}, \citenamefont {Hristov}, \citenamefont {Keating},
  \citenamefont {Lange},\ and\ \citenamefont {Philhour}}]{Leong2002}%
  \BibitemOpen
  \bibfield  {author} {\bibinfo {author} {\bibfnamefont {J.~R.}\ \bibnamefont
  {Leong}}, \bibinfo {author} {\bibfnamefont {R.~S.}\ \bibnamefont {Bhatia}},
  \bibinfo {author} {\bibfnamefont {V.~V.}\ \bibnamefont {Hristov}}, \bibinfo
  {author} {\bibfnamefont {B.~G.}\ \bibnamefont {Keating}}, \bibinfo {author}
  {\bibfnamefont {A.~E.}\ \bibnamefont {Lange}}, \ and\ \bibinfo {author}
  {\bibfnamefont {B.~J.}\ \bibnamefont {Philhour}},\ }\bibfield  {title}
  {\enquote {\bibinfo {title} {Lightweight, space efficient low-pass
  radio-frequency interference filter modules for bolometric detectors},}\
  }\href {\doibase 10.1063/1.1502019} {\bibfield  {journal} {\bibinfo
  {journal} {Review of Scientific Instruments}\ }\textbf {\bibinfo {volume}
  {73}},\ \bibinfo {pages} {3638 -- 3643} (\bibinfo {year} {2002})}\BibitemShut
  {NoStop}%
\bibitem [{\citenamefont {Lukashenko}(2008)}]{Lukashenko2008}%
  \BibitemOpen
  \bibfield  {author} {\bibinfo {author} {\bibfnamefont {A.}~\bibnamefont
  {Lukashenko}, \bibfnamefont {A~.and~Ustinov}},\ }\bibfield  {title} {\enquote
  {\bibinfo {title} {Improved powder filters for qubit measurements},}\ }\href
  {\doibase 10.1063/1.2827515} {\bibfield  {journal} {\bibinfo  {journal}
  {Review of Scientific Instruments}\ }\textbf {\bibinfo {volume} {79}},\
  \bibinfo {eid} {014701} (\bibinfo {year} {2008})}\BibitemShut {NoStop}%
\bibitem [{\citenamefont {Milliken}\ \emph {et~al.}(2007)\citenamefont
  {Milliken}, \citenamefont {Rozen}, \citenamefont {Keefe},\ and\ \citenamefont
  {Koch}}]{Milliken2007}%
  \BibitemOpen
  \bibfield  {author} {\bibinfo {author} {\bibfnamefont {F.~P.}\ \bibnamefont
  {Milliken}}, \bibinfo {author} {\bibfnamefont {J.~R.}\ \bibnamefont {Rozen}},
  \bibinfo {author} {\bibfnamefont {G.~A.}\ \bibnamefont {Keefe}}, \ and\
  \bibinfo {author} {\bibfnamefont {R.~H.}\ \bibnamefont {Koch}},\ }\bibfield
  {title} {\enquote {\bibinfo {title} {50 ohm characteristic impedance low-pass
  metal powder filters},}\ }\href {\doibase 10.1063/1.2431770} {\bibfield
  {journal} {\bibinfo  {journal} {Review of Scientific Instruments}\ }\textbf
  {\bibinfo {volume} {78}},\ \bibinfo {pages} {024701} (\bibinfo {year}
  {2007})}\BibitemShut {NoStop}%
\bibitem [{\citenamefont {Santavicca}\ and\ \citenamefont
  {Prober}(2008)}]{Santavicca2008}%
  \BibitemOpen
  \bibfield  {author} {\bibinfo {author} {\bibfnamefont {D.~F.}\ \bibnamefont
  {Santavicca}}\ and\ \bibinfo {author} {\bibfnamefont {D.~E.}\ \bibnamefont
  {Prober}},\ }\bibfield  {title} {\enquote {\bibinfo {title}
  {Impedance-matched low-pass stripline filters},}\ }\href
  {http://stacks.iop.org/0957-0233/19/i=8/a=087001} {\bibfield  {journal}
  {\bibinfo  {journal} {Measurement Science and Technology}\ }\textbf {\bibinfo
  {volume} {19}},\ \bibinfo {pages} {087001} (\bibinfo {year}
  {2008})}\BibitemShut {NoStop}%
\bibitem [{\citenamefont {Slichter}, \citenamefont {Naaman},\ and\
  \citenamefont {Siddiqi}(2009)}]{Slichter2009}%
  \BibitemOpen
  \bibfield  {author} {\bibinfo {author} {\bibfnamefont {D.~H.}\ \bibnamefont
  {Slichter}}, \bibinfo {author} {\bibfnamefont {O.}~\bibnamefont {Naaman}}, \
  and\ \bibinfo {author} {\bibfnamefont {I.}~\bibnamefont {Siddiqi}},\
  }\bibfield  {title} {\enquote {\bibinfo {title} {Millikelvin thermal and
  electrical performance of lossy transmission line filters},}\ }\href
  {\doibase 10.1063/1.3133362} {\bibfield  {journal} {\bibinfo  {journal}
  {Applied Physics Letters}\ }\textbf {\bibinfo {volume} {94}},\ \bibinfo {eid}
  {192508} (\bibinfo {year} {2009})}\BibitemShut {NoStop}%
\bibitem [{\citenamefont {ITT}(1975)}]{ITT1975}%
  \BibitemOpen
  \bibfield  {author} {\bibinfo {author} {\bibnamefont {ITT}},\ }\href@noop {}
  {\emph {\bibinfo {title} {Reference Data for Radio Engineers}}}\ (\bibinfo
  {publisher} {Howard W. Sams},\ \bibinfo {address} {New York},\ \bibinfo
  {year} {1975})\ Chap.~\bibinfo {chapter} {24}, pp.\ \bibinfo {pages} {21 --
  23}\BibitemShut {NoStop}%
\bibitem [{\citenamefont {Wheeler}(1950)}]{Wheeler1950}%
  \BibitemOpen
  \bibfield  {author} {\bibinfo {author} {\bibfnamefont {H.~A.}\ \bibnamefont
  {Wheeler}},\ }\bibfield  {title} {\enquote {\bibinfo {title}
  {Transmission-line impedance curves},}\ }\href@noop {} {\bibfield  {journal}
  {\bibinfo  {journal} {Proceedings of the IRE}\ }\textbf {\bibinfo {volume}
  {38}},\ \bibinfo {pages} {1400 -- 1403} (\bibinfo {year} {1950})}\BibitemShut
  {NoStop}%
\bibitem [{\citenamefont {Riblet}(1983)}]{Riblet1983}%
  \BibitemOpen
  \bibfield  {author} {\bibinfo {author} {\bibfnamefont {H.~J.}\ \bibnamefont
  {Riblet}},\ }\bibfield  {title} {\enquote {\bibinfo {title} {An accurate
  approximation of the impedance of a circular cylinder concentric with an
  external square tube},}\ }\href@noop {} {\bibfield  {journal} {\bibinfo
  {journal} {Microwave Theory and Techniques, IEEE Transactions on}\ }\textbf
  {\bibinfo {volume} {31}},\ \bibinfo {pages} {841 -- 844} (\bibinfo {year}
  {1983})}\BibitemShut {NoStop}%
\bibitem [{\citenamefont {Moreno}(1944)}]{Moreno1944}%
  \BibitemOpen
  \bibfield  {author} {\bibinfo {author} {\bibfnamefont {T.}~\bibnamefont
  {Moreno}},\ }\href@noop {} {\emph {\bibinfo {title} {Microwave Transmission
  Design Data}}},\ \bibinfo {number} {23 - 80}\ (\bibinfo  {publisher} {Sperry
  Gyroscope Company},\ \bibinfo {address} {New York},\ \bibinfo {year}
  {1944})\BibitemShut {NoStop}%
\bibitem [{\citenamefont {Bohren}\ and\ \citenamefont
  {Huffman}(2004)}]{Bohren2004}%
  \BibitemOpen
  \bibfield  {author} {\bibinfo {author} {\bibfnamefont {C.}~\bibnamefont
  {Bohren}}\ and\ \bibinfo {author} {\bibfnamefont {D.}~\bibnamefont
  {Huffman}},\ }\href@noop {} {\emph {\bibinfo {title} {Absorption and
  Scattering of Light by Small Particles}}},\ \bibinfo {number} {36 - 41}\
  (\bibinfo  {publisher} {Wiley-VCH},\ \bibinfo {address} {Weinheim},\ \bibinfo
  {year} {2004})\BibitemShut {NoStop}%
\bibitem [{\citenamefont {Hipple}(1954)}]{VonHipple1954}%
  \BibitemOpen
  \bibfield  {author} {\bibinfo {author} {\bibfnamefont {A.~V.}\ \bibnamefont
  {Hipple}},\ }\href@noop {} {\emph {\bibinfo {title} {Dielectrics and
  Waves}}}\ (\bibinfo  {publisher} {Wiley},\ \bibinfo {address} {New York},\
  \bibinfo {year} {1954})\ pp.\ \bibinfo {pages} {26 -- 28}\BibitemShut
  {NoStop}%
\bibitem [{\citenamefont {{Wollack}}\ \emph {et~al.}(2008)\citenamefont
  {{Wollack}}, \citenamefont {{Fixsen}}, \citenamefont {{Henry}}, \citenamefont
  {{Kogut}}, \citenamefont {{Limon}},\ and\ \citenamefont
  {{Mirel}}}]{Wollack2008}%
  \BibitemOpen
  \bibfield  {author} {\bibinfo {author} {\bibfnamefont {E.~J.}\ \bibnamefont
  {{Wollack}}}, \bibinfo {author} {\bibfnamefont {D.~J.}\ \bibnamefont
  {{Fixsen}}}, \bibinfo {author} {\bibfnamefont {R.}~\bibnamefont {{Henry}}},
  \bibinfo {author} {\bibfnamefont {A.}~\bibnamefont {{Kogut}}}, \bibinfo
  {author} {\bibfnamefont {M.}~\bibnamefont {{Limon}}}, \ and\ \bibinfo
  {author} {\bibfnamefont {P.}~\bibnamefont {{Mirel}}},\ }\bibfield  {title}
  {\enquote {\bibinfo {title} {{Electromagnetic and Thermal Properties of a
  Conductively Loaded Epoxy}},}\ }\href {\doibase 10.1007/s10762-007-9299-4}
  {\bibfield  {journal} {\bibinfo  {journal} {International Journal of Infrared
  and Millimeter Waves}\ }\textbf {\bibinfo {volume} {29}},\ \bibinfo {pages}
  {51 -- 61} (\bibinfo {year} {2008})}\BibitemShut {NoStop}%
\bibitem [{\citenamefont {Denny}\ and\ \citenamefont
  {Warren}(1968)}]{Denny1968}%
  \BibitemOpen
  \bibfield  {author} {\bibinfo {author} {\bibfnamefont {H.}~\bibnamefont
  {Denny}}\ and\ \bibinfo {author} {\bibfnamefont {W.}~\bibnamefont {Warren}},\
  }\bibfield  {title} {\enquote {\bibinfo {title} {Lossy transmission line
  filters},}\ }\href {\doibase 10.1109/TEMC.1968.302978} {\bibfield  {journal}
  {\bibinfo  {journal} {Electromagnetic Compatibility, IEEE Transactions on}\
  }\textbf {\bibinfo {volume} {EMC-10}},\ \bibinfo {pages} {363 -- 370}
  (\bibinfo {year} {1968})}\BibitemShut {NoStop}%
\bibitem [{\citenamefont {Janezic}\ and\ \citenamefont
  {Jargon}(1999)}]{Janezic1999}%
  \BibitemOpen
  \bibfield  {author} {\bibinfo {author} {\bibfnamefont {M.~D.}\ \bibnamefont
  {Janezic}}\ and\ \bibinfo {author} {\bibfnamefont {J.~A.}\ \bibnamefont
  {Jargon}},\ }\bibfield  {title} {\enquote {\bibinfo {title} {Complex
  permittivity determination from propagation constant measurements},}\ }\href
  {\doibase 10.1109/75.755052} {\bibfield  {journal} {\bibinfo  {journal}
  {Microwave and Guided Wave Letters, IEEE}\ }\textbf {\bibinfo {volume} {9}},\
  \bibinfo {pages} {76 -- 78} (\bibinfo {year} {1999})}\BibitemShut {NoStop}%
\bibitem [{\citenamefont {Pozar}(2004)}]{Pozar2004}%
  \BibitemOpen
  \bibfield  {author} {\bibinfo {author} {\bibfnamefont {D.}~\bibnamefont
  {Pozar}},\ }\href@noop {} {\emph {\bibinfo {title} {Microwave Engineering}}}\
  (\bibinfo  {publisher} {Wiley},\ \bibinfo {address} {New York},\ \bibinfo
  {year} {2004})\BibitemShut {NoStop}%
\bibitem [{\citenamefont {Ogreid}\ and\ \citenamefont
  {Osland}(2001)}]{Ogreid2001}%
  \BibitemOpen
  \bibfield  {author} {\bibinfo {author} {\bibfnamefont {O.~M.}\ \bibnamefont
  {Ogreid}}\ and\ \bibinfo {author} {\bibfnamefont {P.}~\bibnamefont
  {Osland}},\ }\bibfield  {title} {\enquote {\bibinfo {title} {More series
  related to the euler series},}\ }\href {\doibase
  http://dx.doi.org/10.1016/S0377-0427(00)00630-0} {\bibfield  {journal}
  {\bibinfo  {journal} {Journal of Computational and Applied Mathematics}\
  }\textbf {\bibinfo {volume} {136}},\ \bibinfo {pages} {389 -- 403} (\bibinfo
  {year} {2001})}\BibitemShut {NoStop}%
\bibitem [{\citenamefont {Landsberg}\ and\ \citenamefont
  {De~Vos}(1989)}]{Landsberg1989}%
  \BibitemOpen
  \bibfield  {author} {\bibinfo {author} {\bibfnamefont {P.~T.}\ \bibnamefont
  {Landsberg}}\ and\ \bibinfo {author} {\bibfnamefont {A.}~\bibnamefont
  {De~Vos}},\ }\bibfield  {title} {\enquote {\bibinfo {title} {The
  stefan-boltzmann constant in n-dimensional space},}\ }\href
  {http://stacks.iop.org/0305-4470/22/i=8/a=021} {\bibfield  {journal}
  {\bibinfo  {journal} {J. Phys. A: Math. Gen.}\ }\textbf {\bibinfo {volume}
  {22}},\ \bibinfo {pages} {1073 -- 1084} (\bibinfo {year} {1989})}\BibitemShut
  {NoStop}%
\bibitem [{\citenamefont {Menon}\ and\ \citenamefont
  {Agrawal}(1998)}]{Menon1998}%
  \BibitemOpen
  \bibfield  {author} {\bibinfo {author} {\bibfnamefont {V.~J.}\ \bibnamefont
  {Menon}}\ and\ \bibinfo {author} {\bibfnamefont {D.~C.}\ \bibnamefont
  {Agrawal}},\ }\bibfield  {title} {\enquote {\bibinfo {title} {Comment on `the
  stefan-boltzmann constant in n-dimensional space'},}\ }\href
  {http://stacks.iop.org/0305-4470/31/i=3/a=021} {\bibfield  {journal}
  {\bibinfo  {journal} {J. Phys. A: Math. Gen.}\ }\textbf {\bibinfo {volume}
  {31}},\ \bibinfo {pages} {1109} (\bibinfo {year} {1998})}\BibitemShut
  {NoStop}%
\end{thebibliography}
%
%

\end{document}